\documentclass[universe,article,accept,moreauthors,pdftex]{Definitions/mdpi} 
\firstpage{1} 
\pubvolume{1}
\issuenum{1}
\articlenumber{0}
\pubyear{2021}
\copyrightyear{2021}
\externaleditor{Academic Editor: Kirill A. Bronnikov and Sergey Sushkov} 
\datereceived{29 June 2021} 
\dateaccepted{28 July 2021} 
\datepublished{} 
\hreflink{https://doi.org/} 
\usepackage{graphicx}
\usepackage[utf8]{inputenc}
\usepackage{float}
\usepackage{multicol}
\usepackage{placeins}
\usepackage{xcolor}
\usepackage{amsmath,amssymb}

\newcommand{\cl}{\mathcal{L}}
\newcommand{\ce}{\mathcal{E}}
\newcommand{\diff}{\mathrm{d}}

\newcommand{\mnras}{Mon. Not. RAS}
\newcommand{\prd}{Phys. Rev. D}
\newcommand{\jcap}{JCAP}
\newcommand{\prl}{Phys. Rev. Lett.}
\newcommand{\apjl}{Astrophys. J. Lett.}
\newcommand{\apj}{Astrophys. J.}
\newcommand{\aap}{Astron. Astrophys.}

\Title{Epicyclic Oscillations around Simpson--Visser Regular Black Holes and Wormholes} 

\TitleCitation{Epicyclic Oscillations around Simpson--Visser Regular Black Holes and Wormholes}

\Author{Zden\v{e}k Stuchl\'{i}k and Jaroslav Vrba *} 
\AuthorNames{Zden\v{e}k Stuchl\'{i}k and Jaroslav Vrba}

\AuthorCitation{Stuchl\'{i}k, Z.; Vrba, J.}

\address[1]{%
Research Centre of Theoretical Physics and Astrophysics, Institute of Physics, Silesian University in Opava, Bezru\v{c}ovo n\'{a}m. 13, CZ-746 01 Opava, Czech Republic; zdenek.stuchlik@physics.slu.cz\\}
\corres{\hangafter=1 \hangindent=1.05em \hspace{-0.82em}{Correspondence: jaroslav.vrba@physics.slu.cz}}

\abstract{We study epicyclic oscillatory motion along circular geodesics of the Simpson--Visser meta-geometry describing in a unique way regular black-bounce black holes and reflection-symmetric wormholes by using a length parameter $l$. We give the frequencies of the orbital and epicyclic motion in a Keplerian disc with inner edge at the innermost circular geodesic located above the black hole outer horizon or on the our side of the wormhole. We use these frequencies in the epicyclic resonance version of the so-called geodesic models of high-frequency quasi-periodic oscillations (HF QPOs) observed in microquasars and around supermassive black holes in active galactic nuclei to test the ability of this meta-geometry to improve the fitting of HF QPOs observational data from the surrounding of supermassive black holes. We demonstrate that this is really possible for wormholes with sufficiently high length parameter $l$.}


\begin{document}
\maketitle
\section{Introduction}

Simpson and Visser introduced a  very simple theoretically attractive spherically symmetric model of meta-geometry,  coming from the Schwarzschild geometry  and enabling a unique description of regular black holes and wormholes by smooth interpolation between these two possibilities using a length-scale parameter $l$ responsible for regularization of the central singularity and potentially reflecting in a maximally simple way the possible influence of quantum gravity effects; its rotation version has recently also been presented \cite{Sim-Vis:2019:JCAP:,Maz-Fra-Lib:2021:JCAP:}. The Simpson--Visser wormhole is traversable in the same way as the standard Morris--Thorne wormhole solutions \cite{Mor-Tho:1988:AmJPhys:,Mor-Tho-Yur:1988:PhysRevLet:}, thus  representing   a spacetime tunnel connecting distant parts of the Universe (or different universes) that enables the transfer of massive objects. The Simpson--Visser regular black hole demonstrates a black bounce that occurs behind the black hole horizon,  being similar to the idea of the black universe \cite{Bro-Mel-Deh:2007:GReGr:}.
 
The simple case of the reflection-symmetric traversable wormholes was introduced by Visser \cite{Vis:1989:NuclPhysB:,Poi-Vis:1995:PHYSR4:} where two Schwarzschild spacetimes are connected by a spherical shell of extraordinary matter with negative energy density violating the weak energy condition, located at the junction and guaranteeing correctness of the Einstein gravitational equations for the stable traversable wormholes \cite{Vis:1989:PhRvD:}. In the high-dimensional general relativity \cite{Svi-Tah:2018:EPJC:}, or some alternative gravity theories \cite{Har-Lob-Mak:2013:PhRvD:},  such extraordinary form of the stress--energy tensor can be avoided in the wormhole solutions. Moreover, the traversable wormholes constructed without extraordinary forms of matter or alternative gravity are possible for fermions giving a negative Casimir energy \cite{Mal-Mil-Pop:2018:arXiv:}---in the Einstein--Dirac theory  or the Einstein--Maxwell--Dirac theory \cite{Bla-sal-Kno:2020:EPJC:,Bla-sal-Kno-Rad:2021:PhRvL:}. 

Extraordinary recent results of the radio-interferometry observational systems, namely of the Event Horizon Telescope (ETH) and GRAVITY \cite{Tur-Zaj-Eck:2020:ApJ:}, give insight into the innermost region of accretion discs orbiting supermassive black holes such as SgrA* in the Galaxy center, or those in the center of active nucleus of the M87 galaxy. The accreting matter reflecting the shadow of the assumed central rotating Kerr black hole was observed in the SgrA* by GRAVITY, and in the central region of M87 in \cite{EHT:2019:ApJ:} by EHT. These observations of the assumed close vicinity of the black hole horizon inspired a variety of theoretical works representing precision tests of General Relativity in the strongest field limit enabling distinguishing of black hole mimickers of the type of the wormholes from the alternatives as superspinars \cite{Gim-Hor:2009:PhysLetB:,Stu-Hle-Tru:2011:CLAQG:,Stu-Sche:2012:CLAQG:} when the shadow qualitatively (topologically) differs from those corresponding to black holes \cite{Stu-Sche:2010:CLAQG:},  and these astrophysical phenomena are extraordinary even if related to their black hole counterparts \cite{Stu:1980:BAC:,Stu-Sche:2013:CLAQG:,Bla-Stu:2016:PhRvD:,Stu-Bla-Sche:2017:PhRvD:}. 

The optical phenomena related to wormholes were extensively studied for the shadow, weak lensing \cite{Abd-Jur-Ahm-Stu:2016:ASS:}, and strong lensing, but considering the appearance of these phenomena related to processes going on our side of the wormhole. {The epicyclic frequencies in wormhole spacetime were studied by \citet{Del-Kun-Ned:2021:arxiv:}. The optical effects in the Simpson--Visser spacetimes are discussed in \cite{Zho-Xie:2020:EPJC:,Gue-Olm-Rub:2021:arXiv:}.} However, recently, the images of the Keplerian disks located on both sides of the Simpson--Visser wormhole were studied and strong signatures of the optical phenomena phenomena generated by disks on different sides of the wormhole were demonstrated by \citet{Sche-Stu:2021:sub:}. The transition from the regular black hole to the wormhole in the Simpson--Visser meta-geometry state is studied in \cite{Chur-Stu:2020:CLAQG,Bro-Kon:2020:PhRvD:}. 

In the present paper,  we study in the Simpson--Visser meta-geometry the circular geodesic motion and related epicyclic oscillations that are relevant for oscillations of the Keplerian disks. Namely,  we concentrate on the outside of the outer horizon of the regular black holes and on the disk on the our side of the wormhole. We determine frequencies of the orbital motion and the radial and vertical epicyclic oscillations and apply them in the so-called geodesic models of HF QPOs \cite{Stu-Kot-Tor:2013:ASTRA:,Stu-Kol-Kov:2020:UNI:} to test their applicability for explanation of the HF QPOs observed in the microquasars \cite{Tor-etal:2011:ASTRA:}, and especially around supermassive black holes in active galactic nuclei \cite{Smi-Tan-Wag:2021:ApJ}. 

\section{Simpson--Visser Meta-Geometry}

A static and spherically symmetric metric describing the Simpson--Visser meta-geometry governing in a simple way the transition between the regular black-bounce black holes through the null wormhole to the traversable reflection-symmetric wormhole takes in the standard Schwarzschild coordinates the form 
\begin{equation}
	\diff s^2=-f(r)\diff t^2 + \frac{1}{f(r)}\diff r^2 + h(r)\left(\diff\theta^2+\sin^2\theta\,\diff\phi^2\right)
	\label{e:met}
\end{equation}
where  
\begin{eqnarray}
	h(r)&=&r^2+l^2,\\
	f(r)&=&1 - \frac{2M}{\sqrt{h(r)}}
\end{eqnarray}
with $M>0$ giving the ADM mass of the object  and the parameter $l>0$ governing the character of the meta-geometry, guaranteeing the regularization of the central singularity and possibly reflecting the quantum gravity effects. {We have to stress that there is no explicit physical model (given by Lagrangian density) behind the meta-geometry. On the other hand, the geometry regularizes the physical singularity of the standard GR black hole solutions, thus  reflecting  the expected role of quantum gravity having no accepted form at present time. Of course, quantum gravity is not relevant for covering whole the range of the parameter $l/M$ discussed in the present paper. For example, considering the length scale $l=l_\mathrm{Pl}(M/M_\mathrm{Pl})^{1/3}$ corresponding to the Planck density, we find $l/M\sim 10^{-25}(M_\odot/M)^{2/3}$ giving for astrophysically relevant objects very small values that can be related to regular black holes only \cite{Rov-Vid:2014:IJMPD:,Hag-Rov:2015:PRD:}. To obtain values of $l/M>2$ corresponding to wormholes, we have to assume the occurrence of a dynamical process, e.g., of the kind introduced by Malafarina \cite{Mal:2017:Uni:}. In such a case,  there is no reason for $l$ connected to the scale related to quantum gravity, the dynamical process could preserve the horizon keeping the regular black hole state, or destroy the horizon leading to remnant wormhole state \cite{Maz-Fra-Lib:2021:JCAP:}.} Being a trivial modification of the Schwarzschild geometry (corresponding to $l=0$),  the Simpson--Visser geometry may represent three different types of spacetimes. 

For $l<2M$,  it describes a regular black hole where the physical singularity is modified to the so-called black bounce to a different universe through a spacelike throat hidden behind the event horizon; it can   thus also be called hidden wormhole \cite{Car-Rub-Fil-Lib-Vis:2020:PhRvD:}. 

For $l=2M$,  the geometry describes a one-way wormhole with a null throat. {The null wormhole is a one-way wormhole with a null throat $r=0$ being a null surface representing an extremal event horizon.}

For $l>2M$,  the geometry describes a reflection-symmetric traversable (two-way) wormhole of the Morris--Thorne type. In the case of the reflection-symmetric wormholes,  we distinguish two universes---the lower, or   our side of the wormhole ($r>0$), and the upper, or the other side of the wormhole ($r<0$). The wormhole spacetime can be well illustrated by visualization using an embedding diagram of the 2D surface representing, e.g., the constant time section of the equatorial plane to the 3D Euclidean space, as presented in \cite{Sche-Stu:2021:sub:}. 

The Simpson--Visser meta-geometry seems to be very appealing  because of both its simplicity and the unified treatment of distinct kinds of physical objects, namely of the black holes and the wormholes (of course, a wormhole is  also hidden   under the event horizon of the black-bounce regular black holes). It could   thus be  useful in attempts for simple description of the phenomenological models related to the scale of regularization that could reflect,  e.g., quantum gravity effects \cite{Maz-Fra-Lib:2021:JCAP:}. 

\section{Equations of Motion}

We consider the test particle approximation when the motion of a particle having a conserved rest mass $m$ is fully governed by the geodesic structure of the spacetime. The geodesic equations of motion can be found using the standard Hamilton--Jacobi (H-J) method. The H-J action function $S$ fulfills the H-J equation 
\begin{equation}
	-m^2=-\frac{1}{f(r)}\left(\frac{\partial S}{\partial t}\right)^2 + f(r)\left(\frac{\partial S}{\partial r}\right)^2 + \frac{1}{h(r)}\left(\frac{\partial S}{\partial \theta}\right)^2+\frac{1}{h(r)\sin^2\theta}\left(\frac{\partial S}{\partial \phi}\right)^2 . 
\end{equation} 
The action function $S$ is connected to the test particle 4-momentum $p^\mu=\diff x^\mu/\diff\lambda$ by the relation 
\begin{equation}
	p_\mu\equiv \frac{\partial S}{\partial x^\mu}.
\end{equation}

We assume the standard separated form of the action function $S\equiv S_t+S_r+S_\theta+S_\phi$. Due to the spacetime symmetries, the test particle motion must be confined to central planes of the spacetimes and the stationarity of the spacetime implies the constant of motion corresponding to the covariant energy $E\equiv -p_t$, and its axial symmetry implies constant of motion corresponding to the axial angular momentum of the particle $L\equiv p_\phi$. The equations of the geodesic motion can then be given in integrated and separated form 
\begin{eqnarray}
	\left(p^r\right)^2&=&E^2 - f(r)\left(m^2+\frac{L^2+Q}{h(r)}\right),\\
	\left(p^\theta\right)^2&=&\frac{1}{h^2(r)}\left(Q-L^2\cot^2\theta\right),\\
	p^t&=&\frac{E}{f(r)},\\
	p^\phi&=&\frac{L}{h(r)\sin^2\theta},
\end{eqnarray}
where $Q$ is a separation constant and $L^2+Q$ represents the total angular momentum of the particle.
In the case of photons,  the mass parameter takes the value $m=0$. For numerical tractability during the ray-tracing procedure for the photon motion along the null geodesics,  it is convenient to introduce a new latitudinal coordinate $\mu\equiv \cos\theta$ transforming the integrated equations of motion to the form 
\begin{eqnarray}
	\left(p^r\right)^2&=&1 - f(r)\frac{b^2+q}{h(r)},\\
	\left(p^\mu\right)^2&=&\frac{1}{\tilde{h}^2(r)}\left(q-\mu^2(b^2+q)\right),\\
	p^t&=&\frac{1}{f(r)},\\
	p^\phi&=&\frac{l}{\tilde{h}(r)(1-\mu^2)}.
\end{eqnarray}
where the equations of motion are reparameterized by $\lambda\rightarrow E\,\lambda$ and impact parameters are introduced by the relations $b\equiv L/E$ and $q\equiv Q/E^2$. 

For both photons ($m=0$) and test particles ($m>0$), for a single particle, the central plane of the motion can be conveniently chosen as the equatorial plane with $\theta=\pi/2$, $Q=0$, and fixed $p^\theta=0$. The photon motion is studied in   \cite{Sche-Stu:2021:sub:}, which is  devoted to optical phenomena related to wormholes; here, we concentrate on the massive particles, their epicyclic motion around circular orbits, and their possible relation to the observed HF QPOs around microquasars and in active galactic nuclei, taking into account both regular black holes and wormholes. 

\section{Circular Geodesics}

Considering motion of a test particles in the equatorial plane ($\theta=\pi/2$, $\mu=0$), the character of the radial motion can be governed in the standard way \cite{Mis-Tho-Whe:1973:Gravitation:} by the effective potential,  taking for the massive test particles the form 
\begin{equation}
	V^\mathrm{m}_{\mathrm{eff}}(r)=f(r)\left[m^2+\frac{L^2}{h(r)}\right] . 
\end{equation}

\begin{figure} [H]
	\includegraphics[width=\linewidth]{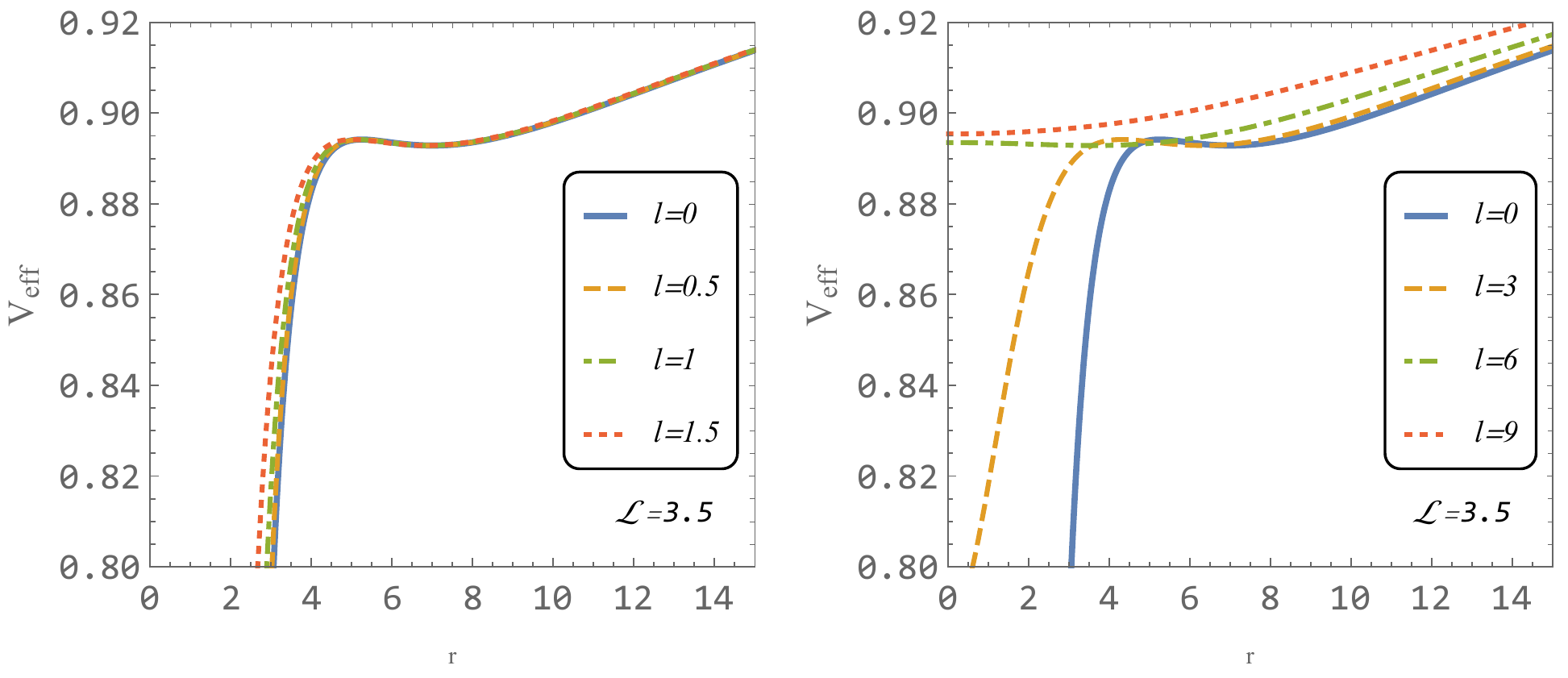}
	\caption{Effective potentials of test particles for characteristic values of the length parameter $l$ and $M=1$. Values of  $l<2M$ giving regular black holes are on the left panel, while values of  $l>2M$ giving wormholes (except the Schwarzschild case $l=0$) are on the right panel.}
	\label{f1:f1}
\end{figure}

The character of the effective potential is illustrated for typical situations characteristic for both   regular black holes and   wormholes in Figure \ref{f1:f1}. The character of the effective potential is similar to the Schwarzschild case for all the regular black holes  and for   wormholes with length parameter $l<6M$. In such spacetimes,  both stable and unstable circular geodesics are possible. The unstable orbits are limited by the photon circular orbit. However, in the wormhole spacetimes with $l\geq 6M$, only the stable circular geodesics are possible.

The circular geodesics are determined by extrema of the effective potential, i.e., by the condition 
\begin{equation}
	\frac{\diff V^\mathrm{m}_{\mathrm{eff}}}{\diff r}=0 
\end{equation}
implying the radial profile of the particle specific angular momentum $\cl= L / m$ in the form 
\begin{equation}
 \cl^{2}_{c}(r,M,l) = \frac{M(r^2+l^2)}{\sqrt{r^2+l^2}-3M}. 
 \label{e:lc}
\end{equation}

The covariant specific energy of the particle on the circular geodesics, $\ce=E/m$, is determined by the value of the effective potential at\ the extremal point, and its radial profile is given by the relation 
\begin{equation}
\ce^{2}_{c}(r,M,l) = \frac{(\sqrt{r^2+l^2}-2M)^2}{\sqrt{r^2+l^2}(\sqrt{r^2+l^2}-3M)}. 
\end{equation}
Note that the specific angular momentum $\cl_{c}$ can be taken with both $+/-$ signs due to the two    (equivalent) possible orientations of the circular motion, while the specific energy $\ce_{c}$ has to be taken with only the positive sign, if we consider the particles in positive root states (for details,  see \cite{Mis-Tho-Whe:1973:Gravitation:,Bic-Stu-Bal:1989:BAC:}). 

\begin{figure} [H]
	\includegraphics[width=\linewidth]{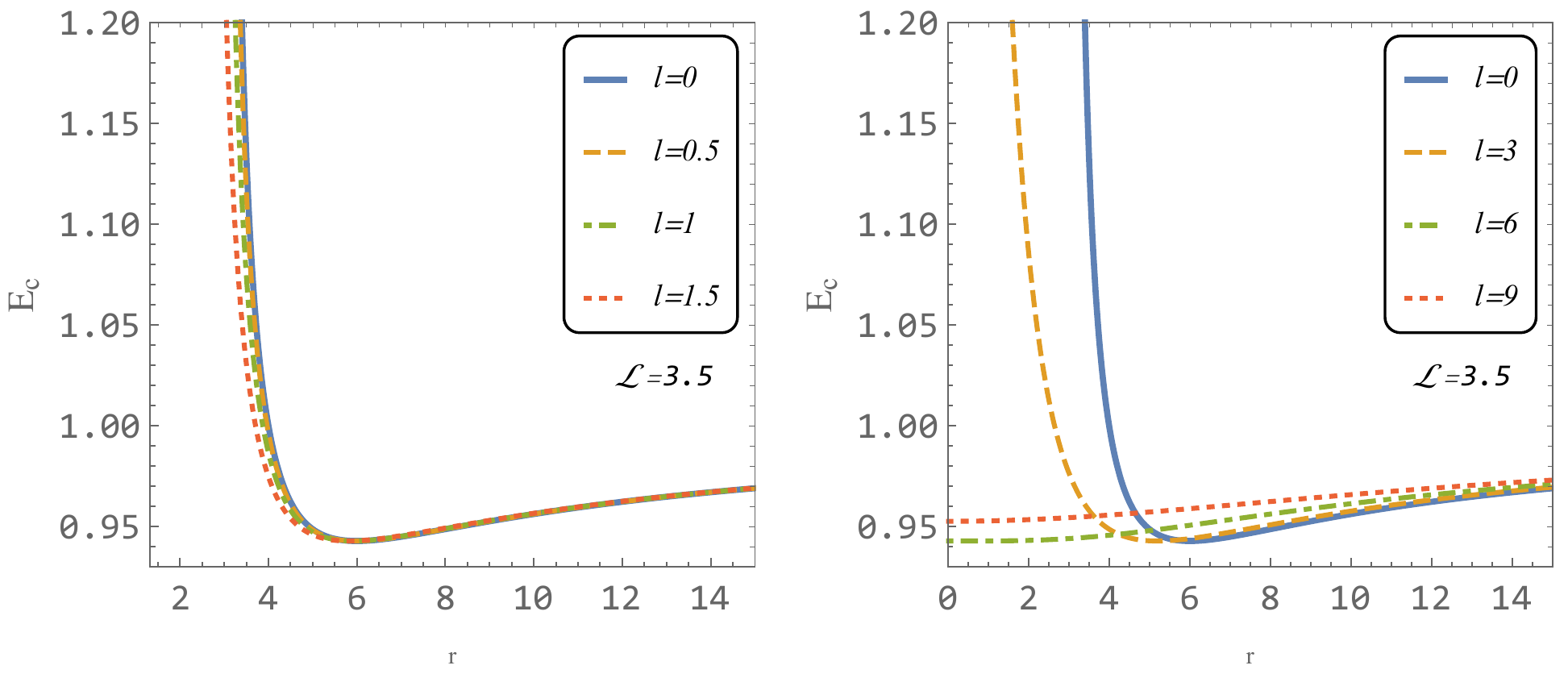}
	\caption{The radial profiles of the specific energy $\ce$ of test particle for characteristic values of the parameter $l$ and $M=1$. Values of  $l<2M$ giving regular black holes are on the left panel, while values of  $l>2M$ giving wormholes (except $l=0$ giving for comparison the Schwarzschild black hole) are on the right panel.}
	\label{f1:f2}
\end{figure}

\begin{figure} [H]
	\includegraphics[width=\linewidth]{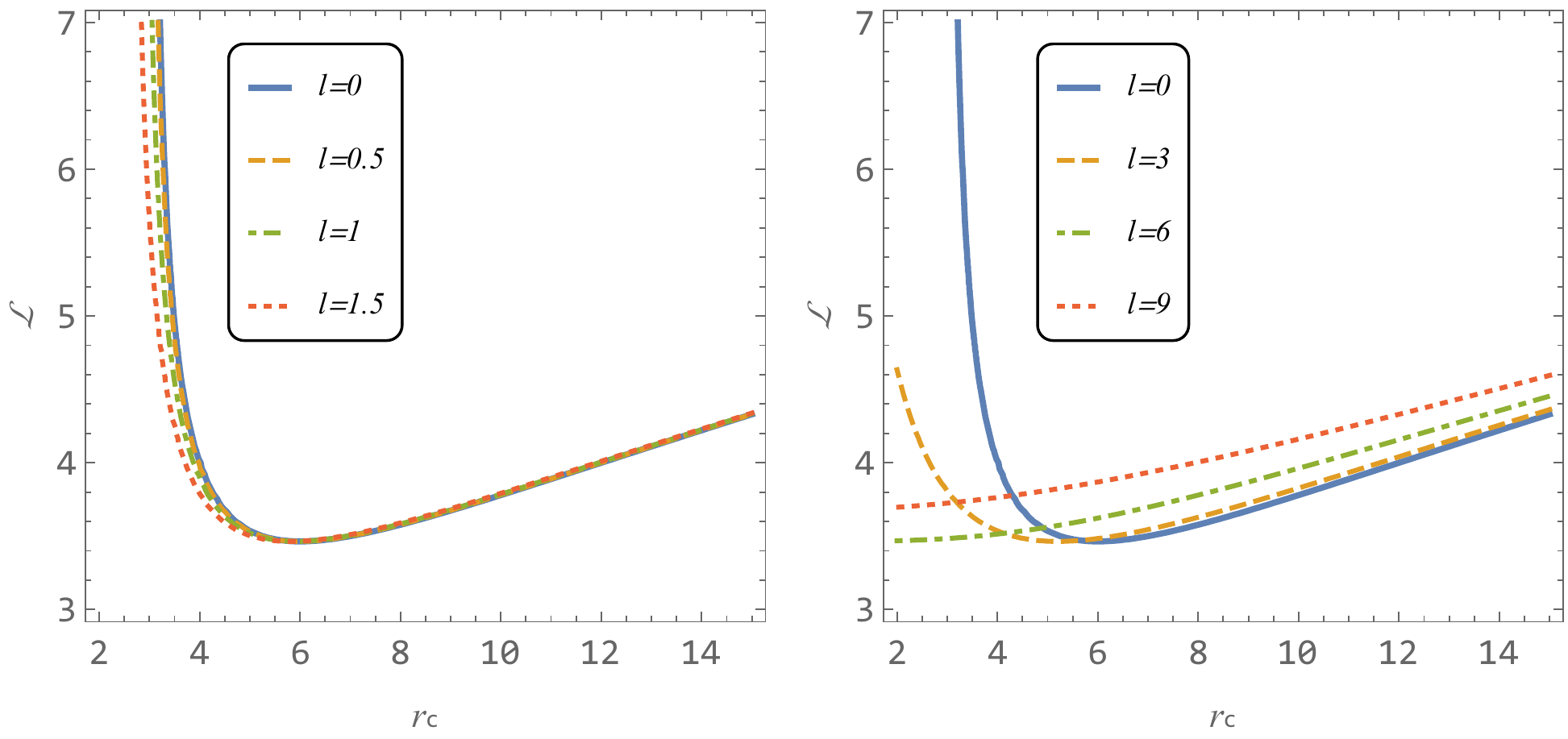}
	\caption{The radial profiles of the specific angular momentum $\cl$ of test particle for characteristic values of the parameter $l$ and $M=1$. Values of  $l$ giving regular black holes are on the left panel, while values of  $l$ giving wormholes (except $l=0$ giving the Schwarzschild  black hole) are on the right panel.}
	\label{f1:f3}
\end{figure}

The radial profiles of the specific energy and specific angular momentum of the circular geodesics are for typical values of $l$ illustrated in Figure \ref{f1:f2} for regular black holes and \mbox{Figure \ref{f1:f3}} for wormholes. Both the specific energy and specific angular momentum radial profiles again demonstrate the fact that for $l>6M$ only stable circular orbits are allowed. 

The stability of the circular geodesic orbits is determined by the sign of $V''_ \mathrm{eff}\equiv\diff^2V_ \mathrm{eff}/\diff r^2$
\begin{equation}
	V''(r_0)
	\left\{\begin{array}{cc}
		<0 & \textrm{stable}\\
		>0 & \textrm{unstable}\\
		=0 & \textrm{marginally stable}
		\end{array}\right.
\end{equation}

The marginally stable (innermost) circular geodesic (ISCO) is determined by the condition 
\begin{equation}
 V''(r_c) = 0 
\end{equation}
that implies the position of the ISCO orbit at the radius 
\begin{equation}
 r_\mathrm{ISCO} = \sqrt{(6M)^2 - l^2} . 
\end{equation}
The corresponding specific angular momentum reads 
\begin{equation}
\cl^2_\mathrm{ISCO} = 12M^2
\end{equation}
being independent of the scale length parameter $l$. 

\textls[+15]{Recall that the position of the event horizon of the regular black holes (with $l<2M$) and the radius of the photon circular geodesic (for $l<3M$) are determined by the relations \cite{Sim-Vis:2019:JCAP:,Sche-Stu:2021:sub:} }\\
\begin{equation}
 r_\mathrm{h} = \sqrt{(2M)^2 - l^2},\quad r_\mathrm{ph} = \sqrt{(3M)^2 - l^2} . 
\end{equation}
The corresponding impact parameter $b=E/L$ of the photon circular orbits is given by 
\begin{equation}
	b^2_\mathrm{ph}=27 M^2
\end{equation}
being independent of the parameter $l$ and having the same form as for photon circular orbit around Schwarzschild black hole.

\begin{figure}[H]
		\includegraphics[scale=1]{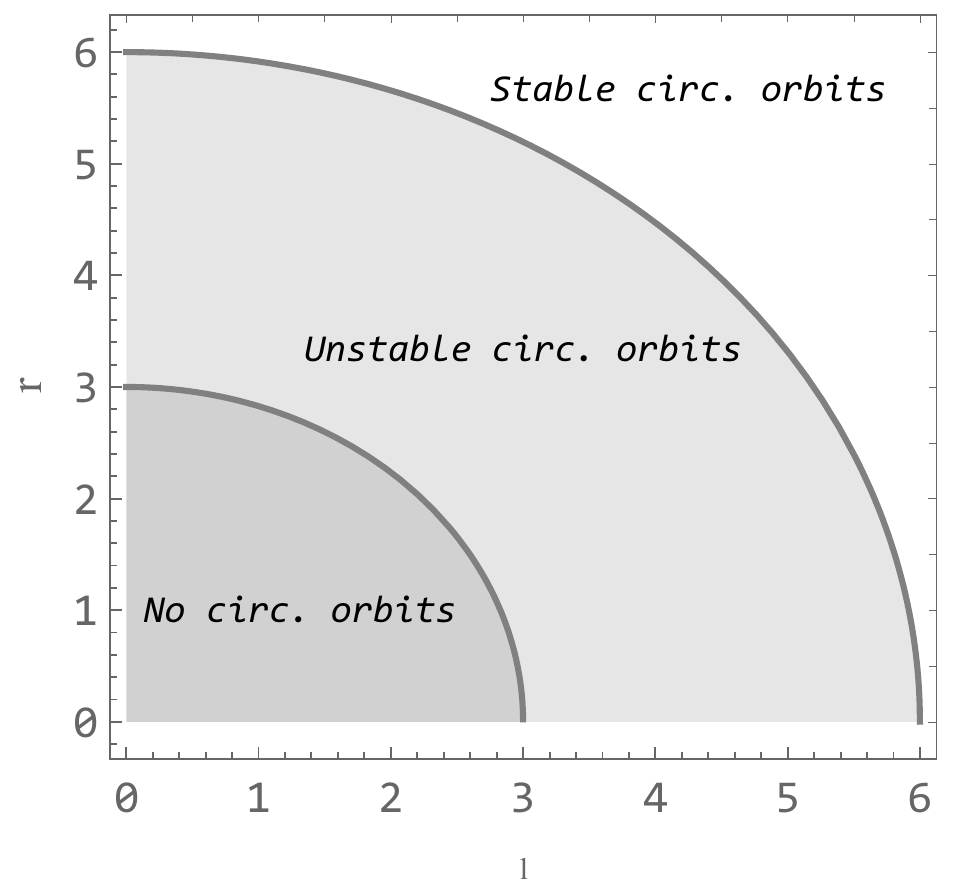}
		\caption{The structure of circular orbits in wormhole spacetime with $M=1$.}
		\label{f1:f4}
\end{figure}

The circular orbits exist where $\ce^2>0$ and $\cl^2>0$. There are two curves, $r_\mathrm{ISCO}$ and $r_\mathrm{ph}$, separating the $r-l$ space into three regions determining existence and position of the photon circular orbits and ISCO (see Figure \ref{f1:f4}). The region where \emph{stable} circular orbits are located is bounded by curve $r_\mathrm{ISCO}(l)$. The \emph{unstable} circular orbits region lays between curves $r_\mathrm{ISCO}$ and $r_\mathrm{ph}$. There are no circular orbits below curve $r_\mathrm{ph}$. 

For completeness,  we also introduce the angular frequency of the geodesic circular motion as related to the static distant observers defined by the relation $\Omega\equiv u^\phi/u^t$---now, the angular frequency, also  called   the Keplerian angular frequency, is given by formula
	\begin{equation}
	\Omega_K^2(r)=\frac{M}{\left(r^2+l^2\right)^{3/2}}.\label{OmegK}
\end{equation}
The angular frequency can be taken with both $+/-$ signs in accordance  with the  orientation of the circular motion. 

Now,  we can study the oscillatory epicyclic motion of test particles orbiting the black hole/wormhole   along a stable circular geodesic assuming only small perturbations from the circular orbits allowing for linear (first-order) perturbation analysis.

\section{\label{s:epmotion}Epicyclic Orbital Motion and Its Frequencies}

A test particle slightly displaced from a stable circular orbit at a radius $r_c$ at the equatorial plane starts oscillatory epicyclic motion around the radius $r_c$ and the latitude $\theta=\pi/2$. We define the coordinate displacement for small perturbations in the radial direction as $r=r_c+\delta r$, and in the latitudinal direction as $\theta=\pi/2+\delta\theta$. The equations governing in the linear perturbation regime the radial and latitudinal epicyclic motion around radius of the stable circular orbit are equivalent to the equation of the harmonic oscillator and take the form \cite{Tor-Stu:2005:ASTRA:,Stu-Kot-Tor:2013:ASTRA:}
\begin{eqnarray}
\delta\ddot{ r}+\bar{\omega}_r^2 \delta r = 0, \ \ \ 
\delta\ddot{ \theta}+\bar{\omega}_\theta^2 \delta\theta = 0.
\end{eqnarray}
where $\bar{\omega_{r}}$ ($\bar{\omega_{\theta}}$) represents the angular velocity of the radial (latitudinal, or vertical) epicyclic oscillations as measured at the radius of the circular orbit.

The orbital angular frequency of the circular motion is determined by the relation 
\begin{equation}
\bar{\omega}_\phi = \frac{\cl}{g_{\theta\theta}}.
\end{equation}

The angular frequencies of the radial and latitudinal epicyclic motion can be easily calculated by using the Hamiltonian formalism \cite{Kol-Stu-Tur:2015:CLAQG:,Tur-Stu-Kol:2016:PHYSR4:,Stu-Kol:2016:EPJC:,Kol-Tur-Stu:2017:EPJC:}. The Hamiltonian generally defined as 
\begin{eqnarray}\label{e:ham}
H=\frac{1}{2}g^{\alpha\beta}p_\alpha p_\beta+\frac{m^2}{2}.
\end{eqnarray}
can be split into its dynamic and potential parts 
\begin{eqnarray}
H=H_{\mathrm{dyn}}+H_{\mathrm{pot}},
\end{eqnarray}
where
\begin{eqnarray}
H_{\mathrm{dyn}}&=& \frac{1}{2}\Big(g^{rr}p_r^2 +g^{\theta\theta}p_\theta^2\Big),\\
H_{\mathrm{pot}}&=& \frac{1}{2}\Big(g^{tt}\mathcal{E}^2+g^{\phi\phi}\cl^2+1\Big).\label{e:hpot}
\end{eqnarray}
The potential part of the Hamiltonian determines the radial and latitudinal epicyclic angular frequencies $\bar{\omega}_r$ and $\bar{\omega}_\theta$ due to the relations 
\begin{eqnarray}
\label{e:bom} 
\bar{\omega}_r^2 &=& \frac{1}{g_{rr}}\frac{\partial^2 H_{\mathrm{pot}}}{\partial r^2},\\ \nonumber
\bar{\omega}_\theta^2 &=& \frac{1}{g_{\theta\theta}}\frac{\partial^2 H_{\mathrm{pot}}}{\partial \theta^2}.
\end{eqnarray}

{Using the relations (\ref{e:met}), (\ref{e:lc}), and (\ref{e:hpot}) in (\ref{e:bom})  and putting for simplicity $M=1$, we arrive at the formula 
\begin{eqnarray}
\label{e:bomegas}
\bar{\omega}_\theta &=& \bar{\omega}_\phi = \frac{1}{\sqrt{\left(l^2+r^2\right) \left(\sqrt{l^2+r^2}-3\right)}},\\ \nonumber
\bar{\omega}_r &=& r\frac{\sqrt{l^2+r^2-3 \left(\sqrt{l^2+r^2}+6\right)}}{\left(l^2+r^2\right)^{5/4} \sqrt{l^2+r^2-9}}.
\end{eqnarray}}
We can see that,  as usual in the spherically symmetric spacetimes, the angular frequency of the latitudinal epicyclic oscillations equals   the angular frequency of the orbital motion. 

Equation (\ref{e:bomegas}) gives  angular frequencies as measured by a local observer, but we have to find the angular frequencies as measured by static observers at infinity who represent real observers who observe the oscillations from large distance. Therefore, we rescale the locally measured angular frequencies by a corresponding redshift factor related to the orbital motion along the stable circular geodesic at $r_c$ and find 
\begin{eqnarray}\label{e:transom}
\omega=\frac{\bar{\omega}}{-g^{tt}\mathcal{E}}. 
\end{eqnarray}

In order to obtain the observed frequencies in the standard units, we have to use the transformation 
\begin{eqnarray}\label{e:transfreq}
\nu=\frac{1}{2\pi}\frac{c^3}{G\,M}\frac{\bar{\omega}}{-g^{tt}\mathcal{E}},
\end{eqnarray}
giving expressions that could be directly used in the fitting to observational data.

\begin{figure} [H]
	\includegraphics[width=.99\linewidth]{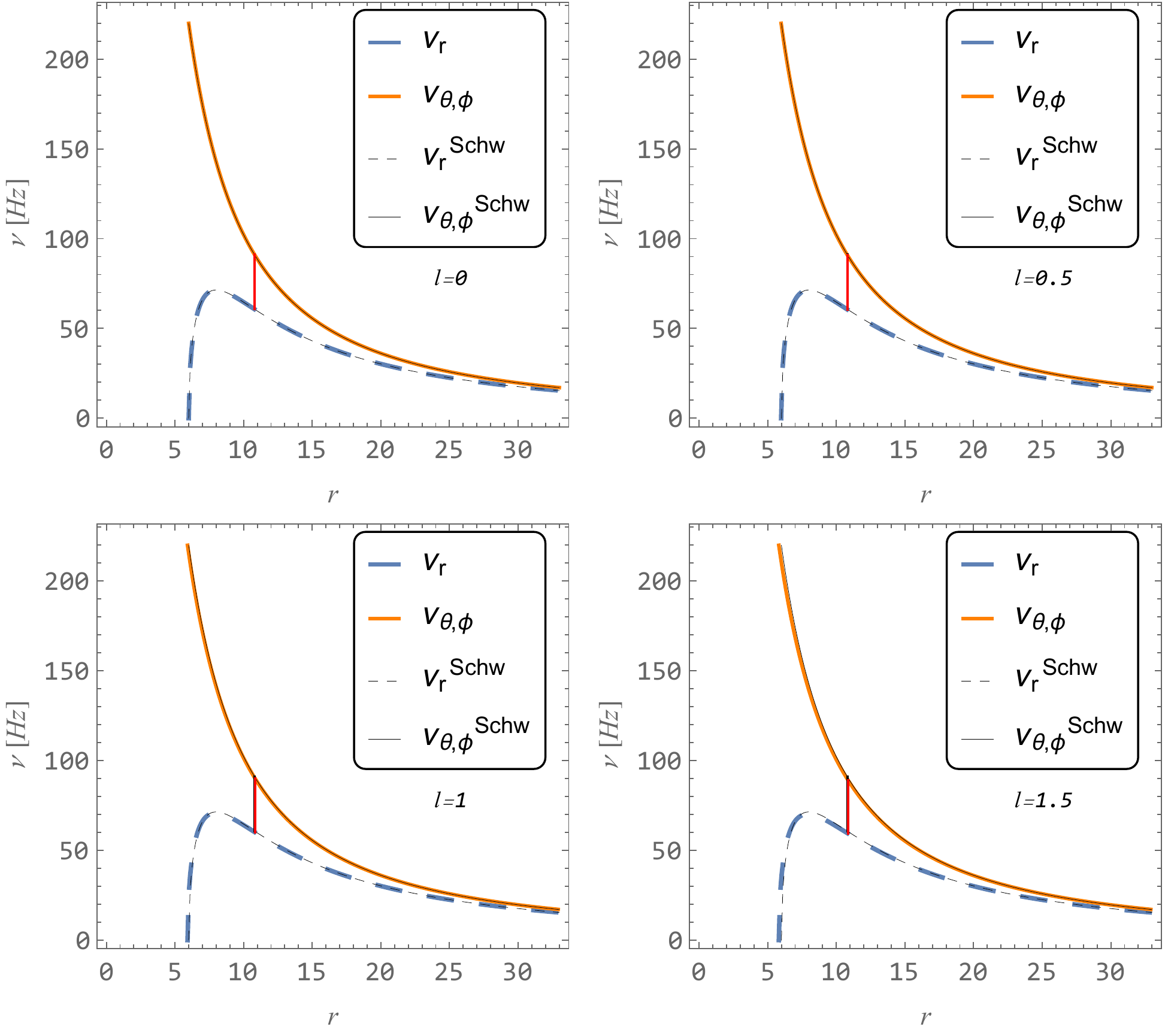}
	\caption{Epicyclic frequencies radial profiles for characteristic values of the parameter $l$ in units of $M$ giving regular black holes and $M=10\,M_\odot$. Vertical black line indicates frequency ratio 3:2 for Schwarzschild case and red for given $l$ (here both vertical lines almost coincide).}
	\label{f1:f5}
\end{figure}

\begin{figure} [H]
	\includegraphics[width=.9\linewidth]{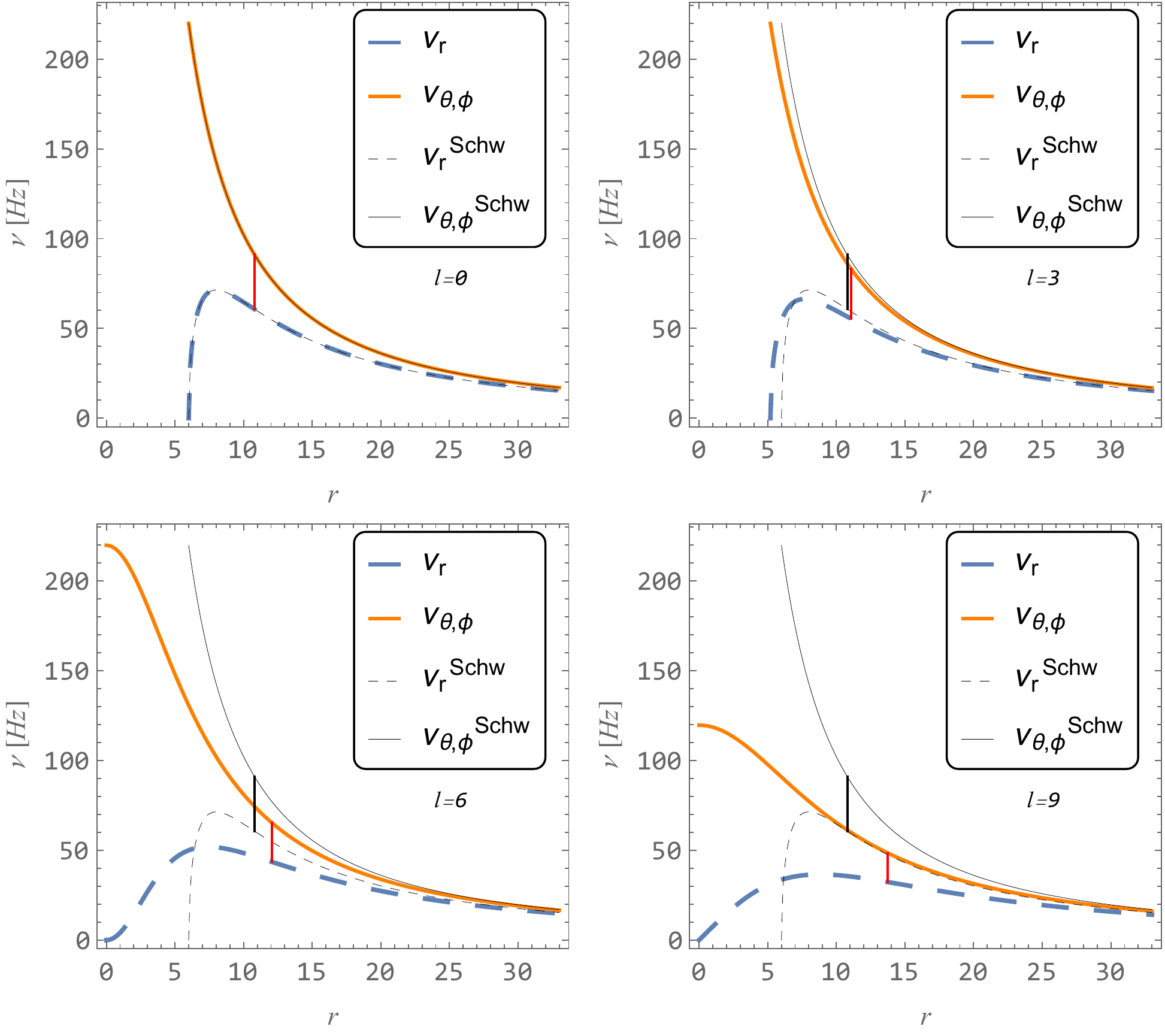}
	\caption{Epicyclic frequencies radial profiles for characteristic values of the parameter $l$ in units of $M$ giving wormholes (except $l=0$ corresponding to Schwarzschild black hole) and $M=10\,M_\odot$. Vertical black line indicates frequencies ratio 3:2 for Schwarzschild case and red for given $l$.}
	\label{f1:f6}
\end{figure}

The radial profiles of the orbital (latitudinal epicyclic) and radial epicyclic frequencies are presented for typical values of the scale length parameter $l$ governing the regular black holes in Figure \ref{f1:f5} and for wormholes in Figure \ref{f1:f6}. We can see that,  in the case of the regular black holes and wormholes with $l<6M$,  the radial profiles of both   frequencies are very similar to those related to the Schwarzschild spacetime, and the frequency ratio 3:2 is located very close to the Schwarzschild position {\cite{Kat-Fuk-Min:1998:BOOK:}}. However, for   wormholes with $l>6M$,  both   epicyclic frequencies radial profiles are significantly modified, being starting at the center $r=0$. The position of the frequency ratio 3:2 is shifted to larger values as compared with the Schwarzschild position, and the shift strongly increases with increasing $l$. 

\begin{figure} [H]
	\includegraphics[scale=0.7]{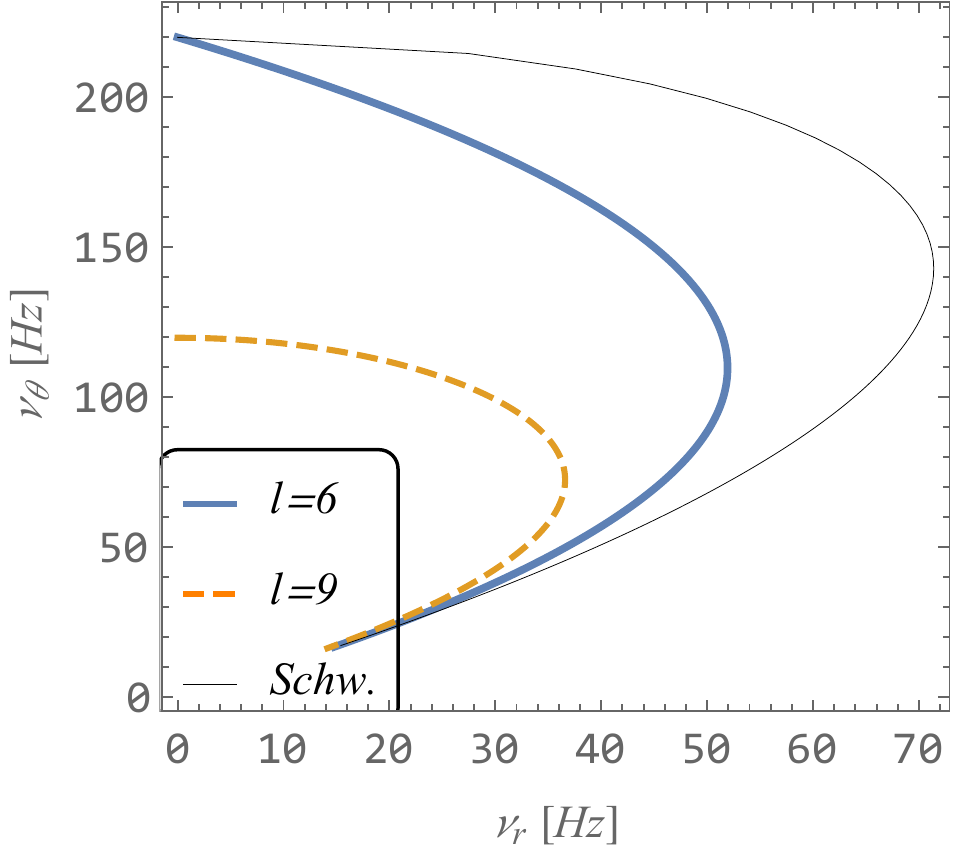}
	\caption{{The dependence of $\nu_\theta$ on $\nu_r$ for the values of the parameter $ l = \{0, \ 6, \ 9 \} $ giving wormholes (except $l=0$ corresponding to the Schwarzschild black hole); $M=10M_\odot$.}}
	\label{f:new}
\end{figure}

{Another view of epicyclic frequencies can be found in Figure \ref{f:new}, where we plot  the dependence of $\nu_\theta$ on $\nu_r$ for the values of the parameter $ l = \{0, \ 6, \ 9 \} $. The values $ l = \{6, \ 9 \} $ were chosen deliberately, as they give the greatest deviation from the result in a purely Schwarzschild background ($ l = 0 $).}
\FloatBarrier
\section{\label{s:astroapp}The Epicyclic Frequencies Applied in the Epicyclic Resonance Model to Fit the Twin HF QPOs with 3:2 Ratio Observed in Microquasars and around Supermassive Black Holes in Active Galactic Nuclei}

In observational astrophysics,  the HF QPOs are of very high importance as they enable well founded predictions of parameters of the black holes in accretion systems, i.e., in microquasars representing binary systems containing a stellar mass black hole, or in active galactic nuclei where supermassive black holes are expected. Frequencies of the HF QPOs are in hundreds of Hz in microquasars, and by six or higher orders (up to 10 orders) smaller around the supermassive black holes. Because of the inverse-mass scaling of the observed frequencies that is typical for the relations governing the epicyclic frequencies of the orbital motion \cite{Rem-McCli:2005:ARAA:}, the models of HF QPOs related to the orbital motion and its epicyclic oscillations, the so-called geodesic models of HF QPOs, are  considered as very promising, especially in connection to the fact that the HF QPOs are often observed in the rational ratio \cite{McCli-Rem:2006:BHbinaries:}, especially in the ratio 3:2 \cite{Tor-etal:2011:ASTRA:}, indicating presence of resonant phenomena \cite{Klu-Abr:2001:ACTAASTR:}. A detailed description of the geodesic models can be found in \cite{Stu-Kot-Tor:2013:ASTRA:}; for generalization with inclusion of an electromagnetic interaction, see the work in  \cite{Kol-Tur-Stu:2017:EPJC:}. In these models,  both   upper $\nu_\mathrm{u}$ and lower $\nu_\mathrm{l}$ observed frequencies are assumed to be a combination of the orbital and epicyclic frequencies (note that these frequencies are also  relevant   for oscillation of slender tori \cite{Rez-Yos-Zan:2003:MNRAS:}). 

The first version of the of the geodesic models is the relativistic precession model proposed in \cite{Ste-Vie:1999:ApJ:} where $\nu_\mathrm{u} = \nu_{\phi}=\nu_\mathrm{K}$ and $\nu_\mathrm{l} = \nu_\mathrm{K} - \nu_\mathrm{r}$. Here,  we consider the simple resonance model discussed in \cite{Tor-Abr-Klu:2005:AA:}, where $\nu_\mathrm{u} = \nu_{\theta}$ and $\nu_\mathrm{l} = \nu_\mathrm{r}$, demonstrating the role of the scale parameter $l$ in the fitting to observational data related to observed 3:2 twin HF QPOs in microquasars  and in HF QPOs observed around supermassive black holes, as discussed recently in \cite{Smi-Tan-Wag:2021:ApJ}, where it is demonstrated that the geodesic models are not able to fit data in the case of   supermassive black holes. We   thus test whether  the proposed scale length factor $l$ that could potentially reflect quantum gravity, or different hidden effects, can play a positive role in fitting the data for   supermassive black holes. 

\begin{figure} [H]
	\includegraphics[width=\linewidth]{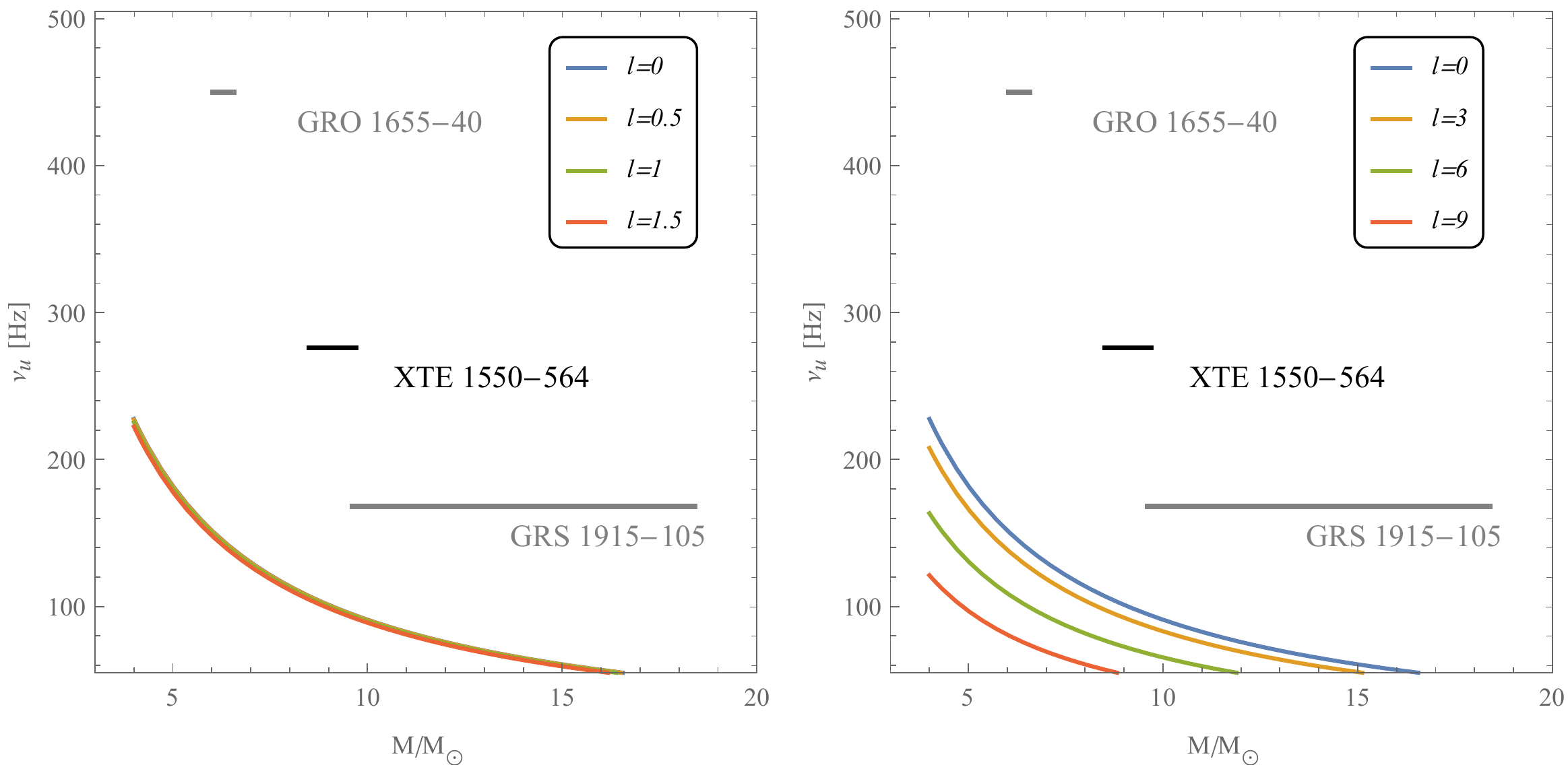}
	\caption{Fit of 3:2 frequencies ratio to mass estimate of microquasars for various parameter $l$ by using the epicyclic resonance model. This model is used in all the following figures of fits. Values of $l$ giving black hole are on the left panel, while values of $l$ giving wormholes (except $l=0$, which gives the Schwarzschild black hole) are on the right panel.}
	\label{f1:f7}
\end{figure}

\begin{figure} [H]
	\includegraphics[width=\linewidth]{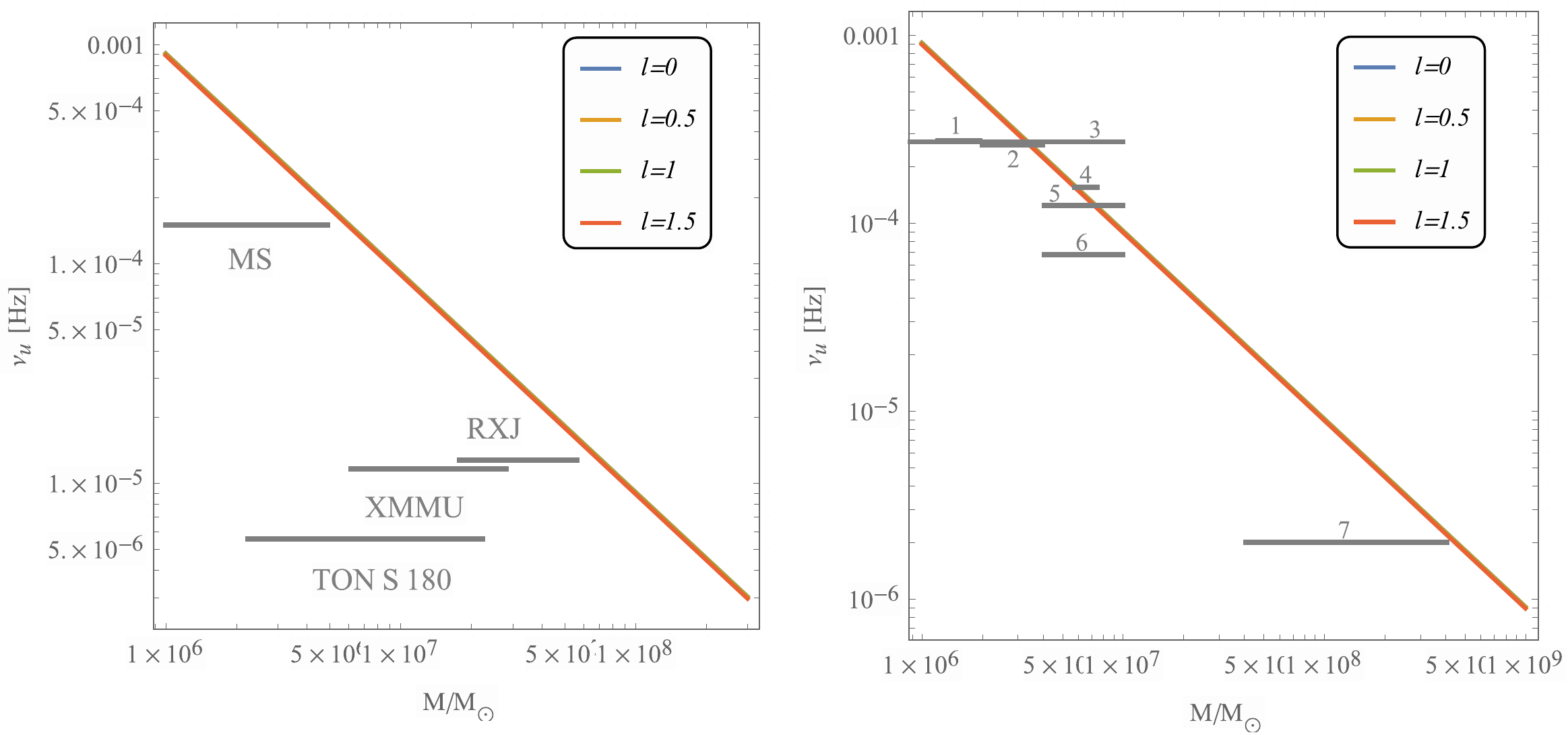}
	\caption{Fit of 3:2 frequencies ratio to mass estimate of assumed supermassive black holes in active galactic nuclei for various values of the parameter $l$ giving regular black holes. The sources with unknown rotation estimates are marked on the left panel. On the right panel are sources with known estimate of spin listed in Table \ref{t:tab1} (cf.  \cite{Smi-Tan-Wag:2021:ApJ}). Notice that the  curves almost coincide.}
	\label{f1:f8}
\end{figure}

\begin{figure} [H]
	\includegraphics[width=\linewidth]{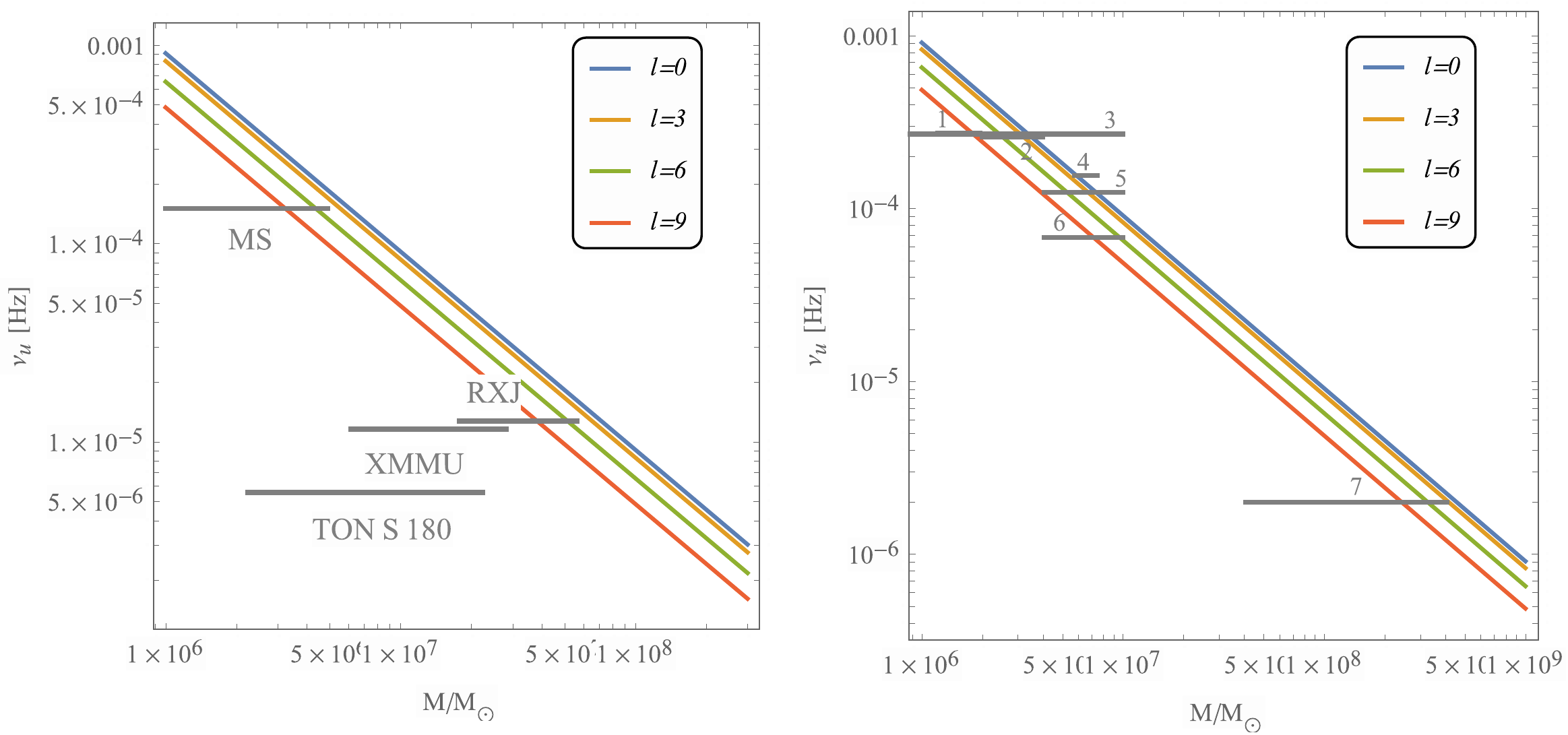}
	\caption{Fit of 3:2 frequencies ratio to mass estimate of assumed supermassive black holes in active galactic nuclei for various values of the parameter $l$ giving Simpson--Visser wormholes. Sources with unknown rotation estimates are marked on the left panel. On the right panel are sources with known estimate of spin listed in Table \ref{t:tab1}   (cf. \cite{Smi-Tan-Wag:2021:ApJ}).}
	\label{f1:f9}
\end{figure}

\begin{figure} [H]
	\includegraphics[width=\linewidth]{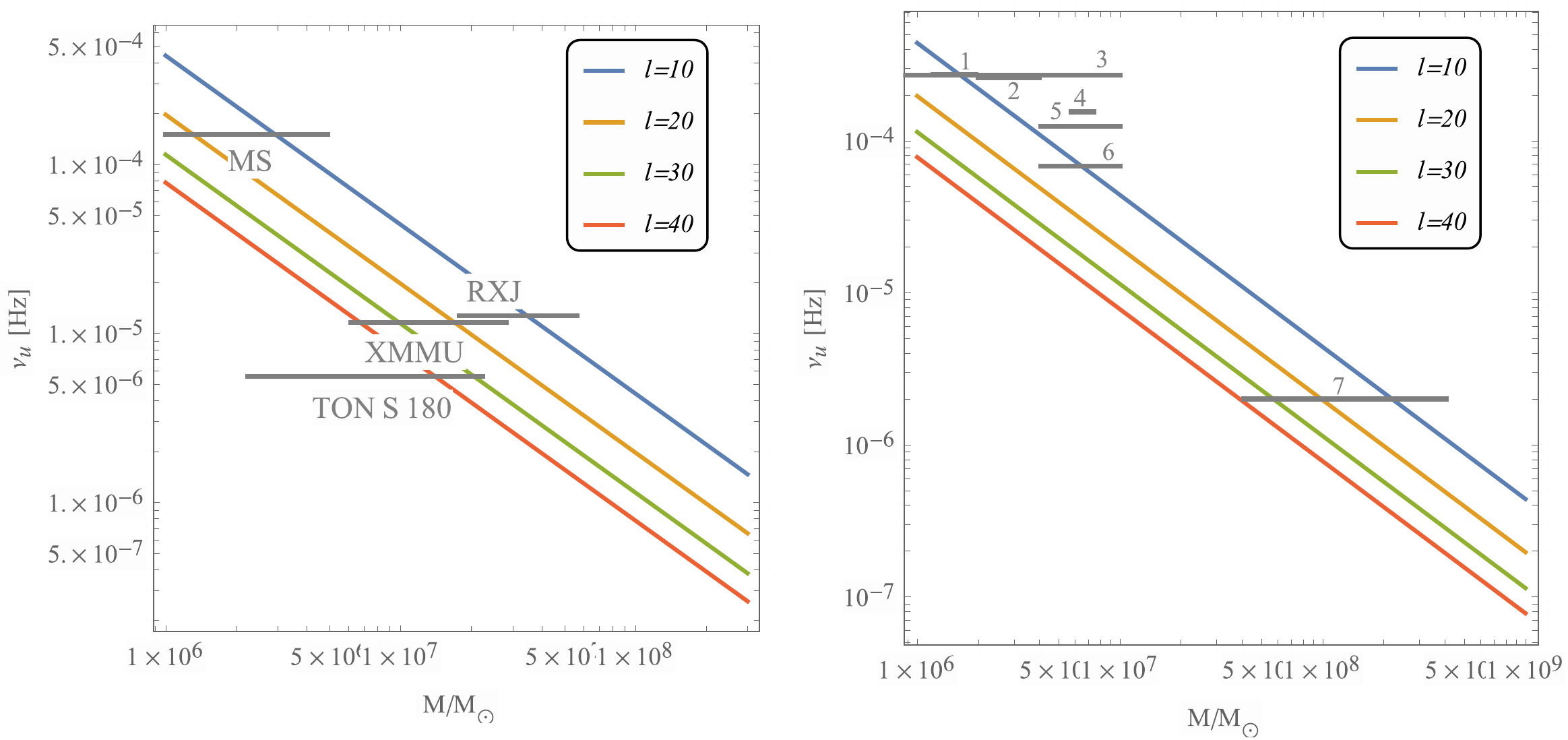}
	\caption{Fit of 3:2 frequencies ratio to mass estimate of assumed supermassive black holes in active galactic nuclei for various values of the parameter $l$ giving wormholes. Sources with unknown rotation estimates are marked in the left panel. On the right panel are sources with a known estimate of spin listed in Table \ref{t:tab1} (cf. \cite{Smi-Tan-Wag:2021:ApJ}).}
	\label{f1:f10}
\end{figure}

In order to fit the data to the model and obtain restrictions on the parameter $l$ from the restriction on the black hole mass obtained by different methods, we use the method developed in \cite{Stu-Kot-Tor:2013:ASTRA:}. We consider all the sources discussed in \cite{Smi-Tan-Wag:2021:ApJ}. The results of the fitting procedure are presented in Figure \ref{f1:f7} for microquasars  and in \mbox{Figures \ref{f1:f8}--\ref{f1:f10}} for the case of supermassive black holes assumed in active galactic nuclei. The results of the fittings in the case of microquasars are always decreasing the possibility to match the observed data. This discrepancy is increasing with increasing of the length parameter $l$. On the other hand, we demonstrate the  possibility to match the data in all the observed sources in active galactic nuclei, assumed to be supermassive black holes---not by regular black holes, but exclusively by wormholes with sufficiently high value of the length parameter $l$, determined in dependence on the concrete source. In some sources,  the values of $l$ could be slightly higher than $l=2M$, but in some other sources they have to be very high, $l \geq 30M$. In the case of large values of $l$, the fitting to the data implies a position of the resonant oscillation in large distance $r>l$ from the center of the wormhole. 
%

\vspace{-6pt}
\begin{specialtable}[H]
\caption{{Table of names of supermassive quasars with estimated spin presented in the right panels of Figures \ref{f1:f8}--\ref{f1:f10} \cite{Smi-Tan-Wag:2021:ApJ}.}}
\label{t:tab1}
\begin{tabular*}{\hsize}{@{}@{\extracolsep{\fill}}ccc@{}}
\toprule
\multicolumn{1}{c}{\textbf{Number}} & \multicolumn{1}{c}{\textbf{Name}} & \multicolumn{1}{c}{\textbf{BH Spin} \boldmath{$a$}} \\ \cmidrule{1-3}
\multicolumn{1}{c}{1} & \multicolumn{1}{c}{MCG-06-30-15} & \multicolumn{1}{c}{>0.917} \\
\multicolumn{1}{c}{2} & \multicolumn{1}{c}{1H0707-495} & \multicolumn{1}{c}{>0.976} \\ 
\multicolumn{1}{c}{3} & \multicolumn{1}{c}{RE J1034+396} & \multicolumn{1}{c}{0.998} \\
\multicolumn{1}{c}{4} & \multicolumn{1}{c}{Mrk 766} & \multicolumn{1}{c}{>0.92} \\ 
\multicolumn{1}{c}{5} & \multicolumn{1}{c}{ESO 113-G010} & \multicolumn{1}{c}{0.998} \\ 
\multicolumn{1}{c}{6} & \multicolumn{1}{c}{ESO 113-G010} & \multicolumn{1}{c}{0.998} \\
\multicolumn{1}{c}{7} & \multicolumn{1}{c}{1H0419-577} & \multicolumn{1}{c}{>0.98} \\ \bottomrule
\end{tabular*}
\end{specialtable}

\vspace{-6pt}

\FloatBarrier

\section{Conclusions}

We   give  frequencies of the epicyclic oscillatory motion around stable circular geodesics in the field of regular black holes and wormholes governed by the Simpson--Visser meta-geometry describing them by a unique scale length parameter that potentially could reflect some hidden effect, e.g., of quantum gravity. The epicyclic frequencies are applied in the special version of the geodesic models of the twin HF QPOs, namely the epicyclic resonance model. We   demonstrate  that,  in the case of   microquasars,  the length parameter $l$ has the  general  tendency to decrease the possibility to find the fits, but,  in the case of the frequencies observed  around assumed supermassive black holes in active galactic nuclei, the parameter acts in positive way, and,  for wormholess with sufficiently high values of $l$,  the fitting is possible for all   observed sources discussed in \cite{Smi-Tan-Wag:2021:ApJ}. Of course, the values of $l$ are very high in many cases, being astrophysically unrealistic, but possible inclusion of the  rotation of the wormholes introduced in \cite{Maz-Fra-Lib:2021:JCAP:} could lead to significant improvement and decrease   the length parameter $l$. {Another positive influence could be due to the dark matter concentrated around the wormhole.}

{Of course, in making definite conclusions, some additional phenomena have to be considered such  as the precession frequencies of some orbiting mass \cite{Riz-Jam-Jus:2019:PRD:}  and the optical phenomena such as shadow \cite{Bam:2013:PRD:,Ned-Tin-Yaz:2013:PRD:}.}

{We can conclude that our positive results for data fitting obtained in the case of supermassive black holes  favor supermassive wormholes in comparison with regular black holes and could be considered as a clear possibility to observationally distinguish    wormholes from black holes---of strong significance from this point of view is the case of large values of $l$ when the oscillations should occur at a rather large distance from the wormhole, as measured in the gravitational radius, contrary to the black hole case when the resonance would occur at a small distance from the black hole. For large values of $l$, we could even expect some influence on the estimation of the supermassive throat object, namely its mass. However, these estimates are usually made for phenomena considered in the weak field limit, at a very large distance from the central object, for $r\lesssim 10^4\,M$, and even for values of $l\lesssim 40M$ we can neglect the role of $l$ at such large distances.}

\vspace{+6pt}
\authorcontributions{Z.S.: writing—original draft preparation and writing—review and editing; J.V.: Sections \ref{s:epmotion} and \ref{s:astroapp}. All authors have read and agreed to the published version of the manuscript.} 

\funding{ZS was supported by the Czech Science Foundation grant No. 19-03950S.}

\acknowledgments{The authors acknowledge the institutional support of the Institute of Physics at the Silesian University in Opava.}

\conflictsofinterest{The authors declare no conflict of interest.}

\end{paracol}
\reftitle{References}


\end{document}